\documentclass[prl,aps,twocolumn,floatfix,superscriptaddress,preprintnumbers]{revtex4-1}
\usepackage{lmodern}
\usepackage{hyperref}
\hypersetup{colorlinks=true,linkcolor=purple,anchorcolor=blue,citecolor=blue, filecolor=blue,urlcolor=red,bookmarksnumbered=true,
pdfview=FitB
}
\usepackage{mathrsfs}
\usepackage{rsfso}
\usepackage{color}
\usepackage{xcolor}
\usepackage{ulem}
\usepackage{csquotes}
\colorlet{purple1}{blue!70!red}
\colorlet{darkred}{red!50!black}

\usepackage{graphicx}
\usepackage[bf,SL,BF]{subfigure}
\usepackage{psfrag}
\usepackage{xcolor}
\usepackage{amssymb}
\usepackage{amsmath}
\usepackage{epstopdf}
\usepackage{natbib}
\newcommand{\be}{\begin{eqnarray}}
\newcommand{\ee}{\end{eqnarray}}

\begin{document}

\title{Nucleon spin decomposition with one dynamical gluon}

\author{Siqi~Xu}
\email{xsq234@163.com} 
\affiliation{Institute of Modern Physics, Chinese Academy of Sciences, Lanzhou 730000, China}
\affiliation{School of Nuclear Science and Technology, University of Chinese Academy of Sciences, Beijing 100049, China}

\author{Chandan~Mondal}
\email{mondal@impcas.ac.cn} 
\affiliation{Institute of Modern Physics, Chinese Academy of Sciences, Lanzhou 730000, China}
\affiliation{School of Nuclear Science and Technology, University of Chinese Academy of Sciences, Beijing 100049, China}

\author{Xingbo~Zhao}
\email{xbzhao@impcas.ac.cn} 
\affiliation{Institute of Modern Physics, Chinese Academy of Sciences, Lanzhou 730000, China}
\affiliation{School of Nuclear Science and Technology, University of Chinese Academy of Sciences, Beijing 100049, China}

\author{Yang~Li}
\email{leeyoung1987@ustc.edu.cn} 
\affiliation{Department of Modern Physics, University of Science and Technology of China, Hefei 230026, China}

\author{James~P.~Vary}
\email{jvary@iastate.edu} 
\affiliation{Department of Physics and Astronomy, Iowa State University, Ames, Iowa 50011, USA}

\collaboration{BLFQ Collaboration}

\date{\today}

\begin{abstract}
We solve for the light-front wave functions of the nucleon from a light-front quantum chromodynamics (QCD) effective Hamiltonian with three-dimensional confinement. We obtain solutions using constituent three quarks combined with three quarks and one gluon Fock components. The resulting light-front wave functions provide a good quality description of the nucleon's quark distribution functions following QCD scale evolution. We present the effects from incorporating a dynamical gluon on the nucleon's gluon densities, helicity distribution and orbital angular momentum that constitutes the nucleon spin sum rule.
\end{abstract}

\maketitle

{\it Introduction.}---QCD is the accepted theory for the strong interactions~\cite{Callan:1977gz}, where nucleons are deemed as confined systems of quarks and gluons, together known as partons. However, it is not yet possible to predict, directly from QCD, the nucleon's global static properties: mass, spin, size etc. This is due to insufficient knowledge of the nonperturbative aspects of QCD that account for colour confinement and chiral symmetry breaking.  Successful theoretical frameworks for predicting some aspects of mass spectra and revealing partonic structures are the Dyson-Schwinger equations (DSEs) of QCD~\cite{Maris:2003vk,Roberts:1994dr,Bashir:2012fs} and discretized space-time Euclidean lattice~\cite{Joo:2019byq,Bazavov:2009bb,Durr:2008zz,Hagler:2009ni}. Complementing these, the Hamiltonian formulation of QCD quantized on the light front (LF)~\cite{Brodsky:1997de,Bakker:2013cea} has advanced in a number of directions. Further enlightenment of nonperturbative QCD can be gained from LF holography~\cite{Brodsky:2014yha,deTeramond:2005su,deTeramond:2008ht}. Meanwhile, basis light-front quantization (BLFQ), which is that approach adopted here, provides a nonperturbative framework for solving relativistic many-body bound state problems in quantum field theories~\cite{Vary:2009gt,Zhao:2014xaa,Nair:2022evk,Wiecki:2014ola,Li:2015zda,Jia:2018ary,Lan:2019vui,Tang:2019gvn,Xu:2019xhk,Xu:2021wwj,Kuang:2022vdy,Lan:2021wok}. 

We adopt an effective LF Hamiltonian and solve for its mass eigenstates at the scales suitable for low-resolution probes within the framework of BLFQ~\cite{Vary:2009gt}. With quark ($q$) and gluon ($g$) being the explicit degrees of freedom for the strong interaction, the Hamiltonian includes LF QCD interactions~\cite{Brodsky:1997de} relevant to constituent $|qqq\rangle$ and $|qqqg\rangle$ Fock sectors of the baryons with a complementary three-dimensional confinement~\cite{Li:2015zda}. 
After fitting Hamiltonian parameters to mass spectra, we compute the nucleon parton distribution functions (PDFs) from the wave functions attained as eigenvectors of the Hamiltonian. The PDFs encoding the nonpertubative structure of the nucleon in terms of the number densities of its confined
constituents, are functions of the longitudinal
momentum fraction ($x$) of the nucleon carried by the
constituents. At leading twist, the complete spin structure
of the nucleon is characterized in terms of three independent
PDFs, namely, unpolarized, helicity,
and transversity.
With our truncated Fock space, we interpret our model as appropriate to a low energy scale and we employ QCD evolution of the PDFs to higher momentum scales to compare results with global analyses.

Two salient issues can be addressed with our approach. The first issue is the gluon density at a low energy scale that contributes to all the parton densities (sea and gluon) under QCD scale evolution. 
The second issue concerns the description of experimental data on the gluon helicity contribution $\Delta G$ to the nucleon spin sum rule: 
$\frac{1}{2}=\frac{1}{2}\Delta \Sigma + \Delta G + L_q + L_g$, with quark helicity $\frac{1}{2}\Delta \Sigma$, quark orbital angular momentum (OAM) $L_q$, and gluon OAM $L_g$ contributions. 
 The RHIC spin program at BNL has revealed that $\Delta G$ is nonvanishing and likely sizable~\cite{STAR:2014wox,deFlorian:2014yva,Nocera:2014gqa,Ethier:2017zbq,STAR:2021mqa}. Together with the known quark helicity contribution $\Delta\Sigma\sim 30\%$, the result  manifests that the other three terms provide a significant fraction of the nucleon spin. Yet, there remain large uncertainties about the small-$x$ contribution to $\Delta G$ defined as the first moment of the polarized gluon PDF: $\Delta G=\int_0^1 dx \Delta G(x)$. For a recent review, see~\cite{Ji:2020ena}. Resolving this matter is one of the prime goals of the future EIC~\cite{AbdulKhalek:2021gbh}. Similar to the scale dependence of the angular momentum
observables~\cite{Thomas:2008ga,Jia:2012wf}, addressing these two issues demands a unified framework, such as we demonstrate here. In particular, we address the first issue by encapsulating properties of the nucleon
at its model scale. Then, we address the second issue by applying QCD evolution to compare with available data
across various other scales where we obtain agreement within sensible precision.

{\it Nucleon wave functions from light-front QCD Hamiltonian.}---The structural information of a bound state is encoded in the light-front wave functions (LFWFs), which are obtained by
solving the eigenvalue problem of the Hamiltonian: $P^-P^+|{\Psi}\rangle=M^2|{\Psi}\rangle$, where $P^\pm=P^0 \pm P^3$ defines the longitudinal momentum ($P^+$) and the LF Hamiltonian ($P^-$) of the system, respectively, with $M^2$ being the mass squared eigenvalue. At fixed LF time ($x^+=t+z=0$), the nucleon state can be expressed as 
\begin{align}\label{Eq1}
|\Psi\rangle=\psi_{(qqq)}|qqq\rangle+\psi_{(qqqg)}|qqqg\rangle + \dots\, , 
\end{align}
where the $\psi_{(\dots)}$ describe the probability amplitudes to attain different parton configurations in the nucleon. These amplitudes can be used to define the LFWFs either in coordinate or momentum space.

At the initial scale, where the baryons are defined in terms of $|qqq\rangle$ and $|qqqg\rangle$ components, we consider the LF Hamiltonian $P^-= P^-_{\rm QCD} +P^-_C$, where $P^-_{\rm QCD}$ and $P^-_{C}$ are respectively the LF QCD Hamiltonian that incorporates interactions relevant to those leading two Fock components and a model for the confining interaction.
In LF gauge, with one dynamical gluon~\cite{Brodsky:1997de}
\begin{align}
P_{\rm QCD}^-=& \int d^2 x^{\perp}dx^- \Big\{\frac{1}{2}\bar{\psi}\gamma^+\frac{m_{0}^2+(i\partial^\perp)^2}{i\partial^+}\psi\nonumber\\
& -\frac{1}{2}A_a^i\left[m_g^2+(i\partial^\perp)^2\right] A^i_a +g_s\bar{\psi}\gamma_{\mu}T^aA_a^{\mu}\psi \nonumber\\
&+\frac{1}{2}g_s^2\bar{\psi}\gamma^+T^a\psi\frac{1}{(i\partial^+)^2}\bar{\psi}\gamma^+T^a\psi \Big\}\,,\label{eqn:PQCD}
\end{align}
where $\psi$ and $A^\mu$ represent the quark and gluon fields, respectively. $T$ is the generator of the $SU(3)$ gauge group in color space, and $\gamma^\mu$ are the Dirac matrices. The first and second terms in Eq.~\eqref{eqn:PQCD} are the kinetic energies of the quark and gluon with bare mass $m_0$  and $m_g$. While the gluon mass is zero in QCD, we permit a phenomenological gluon mass to fit the nucleon form factors (FFs) in our model. The last two terms are the vertex and instantaneous interactions with coupling $g_s$. Following a Fock sector-dependent renormalization procedure developed for positronium in a basis embodying $|e\bar{e}\rangle$ and $|e\bar{e}\gamma\rangle$~\cite{Zhao:2014hpa,Zhao:2020kuf} and further employed for mesons~\cite{Lan:2021wok}, we produce the quark mass counter term ($\delta m$) and define $m_0=m_q+\delta m$, where $m_{q}$ is the physical quark mass. In this initial work, we neglect antisymmetrization of identical quarks. Referring to Ref.~\cite{Glazek:1992aq}, we allow an independent quark mass $m_f$ in the vertex interaction. 

We consider confinement in the leading Fock sector~\cite{Li:2015zda}
\begin{equation}
P_{\rm C}^-P^+ =\frac{\kappa^4}{2}\sum_{i\ne j} \left\{\{\vec{r}_{ij\perp}^{~2} -\frac{\partial_{x_i}(x_i x_j\partial_{x_j})}{(m_i+m_j)^2}\right\}\,, \label{eqn:PC}
\end{equation}
where $\vec{r}_{ij\perp}=\sqrt{x_{i} x_{j}}(\vec{r}_{i\perp}-\vec{r}_{j\perp})$ is the relative coordinate related to the holographic variable~\cite{Brodsky:2014yha}, $\kappa$ is the strength of the confinement and $\partial_{x}\equiv (\partial/\partial x)_{r_{ij\perp}}$. 
We omitted explicit confinement in the $|qqqg\rangle$ sector with the hope that the limited basis in the transverse direction and the massive gluon retain adequate effects of confinement at this stage in the development. Notably, massive gluons appear in functional approaches such as DSEs approach as a consequence of confinement effects \cite{Alkofer:2000wg}.

We solve the eigenvalue problem of the LF Hamiltonian with the framework of BLFQ~\cite{Vary:2009gt}. We employ a plane-wave state in the longitudinal direction confined in a one dimensional box of length $2L$ with antiperiodic (periodic)  boundary conditions for quarks (gluons), two-dimensional harmonic oscillator (``2D-HO") wave function, $\Phi_{nm}(\vec{p}_\perp;b)$ with scale parameter $b$, in the transverse direction, and light-cone helicity state in the spin space~\cite{Zhao:2014xaa} to transform the eigenvalue problem of the Hamiltonian to a hermitian matrix eigenvalue problem. The above basis choice introduces four quantum numbers for each parton single-particle state, $\bar{\alpha}=\{k,n,m,\lambda\}$. Here, $ k $ represents the longitudinal degree of freedom that corresponds to the parton longitudinal momentum $p^{+}=\frac{2\pi k}{L} $ with $ k $ taking positive half-integer (integer) values for quarks (gluons). We omit the zero mode for gluons. The 2D-HO wave function carries the principal and orbital angular quantum number denoted by $n$ and $m$, respectively  and $ \lambda $ represents the spin. In the case of Fock sectors allowing for multiple color-singlet state, we need an additional label to identify each color singlet state. Note that $|qqqg\rangle$ has two color-singlet states. 

Two basis space truncations, $ N_{\rm max} $ and $K$, are introduced to render the resulting matrix finite~\cite{Zhao:2014xaa}. $ N_{\rm max} $ acts as truncation in the transverse direction for the total energy of the 2D-HO basis states: $ \sum_i \left( 2 n_i +|m_i| +1  \right) \leq N_{\rm max} $, and $\sum_i k_i =K$. The longitudinal momentum fraction is then expressed as $x_i=k_i/K$. $K$ represents the resolution in the longitudinal direction and hence a resolution for the PDFs.

The resulting LFWFs of the baryons with helicity $\Lambda$ in momentum space are then expressed in terms of components within each of the Fock sectors as 
\begin{equation}
\Psi^{\mathcal{N},\,\Lambda}_{\{x_i,\vec{p}_{\perp i},\lambda_i\}} =\sum_{ \{n_i m_i\} }\psi^{\mathcal{N}}({\{\overline{\alpha}_i}\})\prod_{i=1}^{\mathcal{N}}  \phi_{n_i m_i}(\vec{p}_{\perp i},b)\,,
\label{eqn:wf}
\end{equation}
where $\psi^{\mathcal{N}=3}(\{\overline{\alpha}_i\})$ and $\psi^{\mathcal{N}=4}(\{\overline{\alpha}_i\})$ are the components of the eigenvectors associated with the Fock sectors $|qqq\rangle$ and $|qqqg\rangle$, respectively, in the BLFQ basis implemented for diagonalizing the full Hamiltonian matrix.

 All the calculations are performed with $\{N_{\rm max},K\}\,$=$\,\{9,16.5\}$. We select the HO scale parameter $b=0.7$ GeV, the UV cutoff for the instantaneous interaction $b_{\rm inst}=3$ GeV, and set our model parameters $\{m_u,m_d, m_g, \kappa, m_f, g_s  \}= \{0.31,0.25, 0.50, 0.54,1.80, 2.40\}$ (all are in units of GeV except $g_s$) by fitting the proton mass~($M$), electromagnetic properties, and its flavor FFs~\cite{Cates:2011pz,Qattan:2012zf,Diehl:2013xca}. 

We compute the electromagnetic radii of the proton from the slope of the Sachs FFs~\cite{Ernst:1960zza} and find its charge radius $\sqrt{\langle r^2_{E}\rangle}=0.85\pm 0.01$ fm and the magnetic radius $\sqrt{\langle r^2_{M}\rangle}=0.88\pm 0.07$ fm, which agree with experimentally measured data $\sqrt{\langle r^2_{E}\rangle_{\rm exp}}=0.840^{+0.003}_{-0.002}$ fm and $\sqrt{\langle r^2_{M}\rangle_{\rm exp}}=0.849^{+0.003}_{-0.003}$ fm~\cite{Lin:2021xrc}.
%
We obtain the magnetic moment of  the proton, $\mu_{\rm{p}}=2.443\pm0.027$, close to the recent lattice QCD simulations: $\mu_{\rm{p}}^{\rm lat}=2.43(9)$~\cite{Alexandrou:2018sjm}, while the experimental value of the magnetic moment is $\mu_{\rm{p}}^{\rm exp}=2.79$~\cite{Tanabashi:2018oca}. 

{\it Parton distribution functions.}---With our resulting LFWFs, the proton's valence quarks and gluon unpolarized and helicity PDFs are given by
\begin{align}
&f(x)= \int_{\mathcal{N}}  \Psi^{\mathcal{N},\,\Lambda\,*}_{\{x_i,\vec{p}_{\perp i},\lambda_i\}}\,\Psi^{\mathcal{N},\,\Lambda}_{\{x_i,\vec{p}_{\perp i},\lambda_i\}}\,\delta(x-x_i)\,,\\
&\Delta f(x)= \int_{\mathcal{N}} \lambda_1\, \Psi^{\mathcal{N},\,\Lambda\,*}_{\{x_i,\vec{p}_{\perp i},\lambda_i\}}\,\Psi^{\mathcal{N},\,\Lambda}_{\{x_i,\vec{p}_{\perp i},\lambda_i\}}\,\delta(x-x_i)\,,
\label{eqn:pdf_i}
\end{align}
respectively, with $f\equiv q,\,g$. We use the abbreviation $\int_{\mathcal{N}} \equiv  \sum_{\mathcal{N},\,\lambda_i}\prod_{i=1}^\mathcal{N} \int \left[\frac{dx\,d^2\vec{p}_\perp}{16\pi^3}\right]_i16\pi^3$ $  \delta(1-\sum x_j)$ $\delta^2(\sum \vec{p}_{\perp j})$ and $i=q,\,g$ labels the valence quarks and gluon, respectively and $\lambda_1=1\,(-1)$ for the struck parton helicity. At our model scale the PDFs for the valence quarks are normalized as $\int_0^1 q (x)\,dx=n_q$, with $n_q$ being the number of quarks of flavor $q$ in the proton and those PDFs together with the gluon PDF complete the momentum sum rule: 
$\int_0^1 \sum_i x f^i(x) dx=1$.

\begin{figure}
\begin{center}
\includegraphics[width=0.95\linewidth]{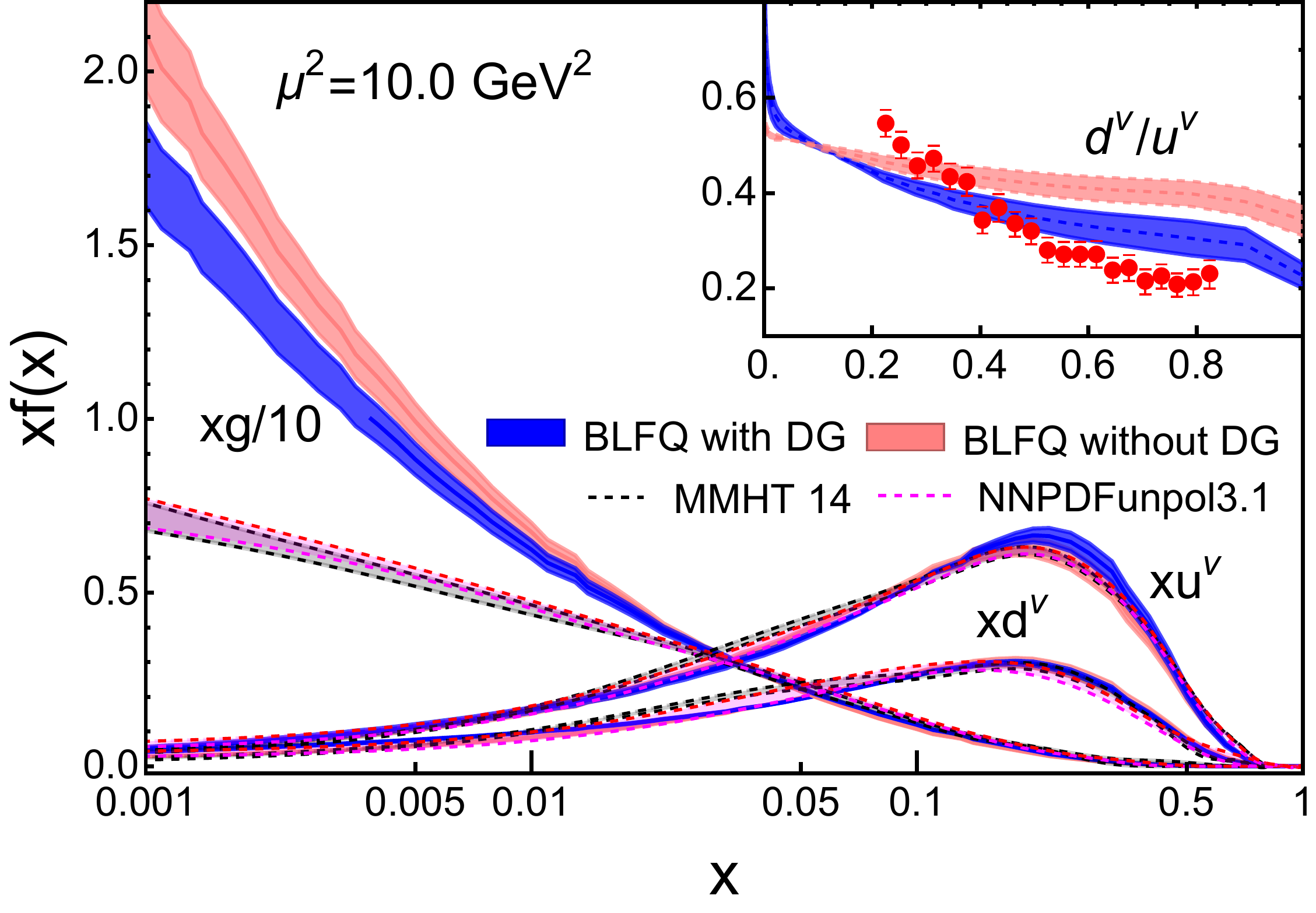}
\caption{the unpolarized valence quark and gluon PDFs of the proton. Our results (blue bands) obtained with one dynamical gluon are compared with the proton PDFs (pink bands) previously obtained from a LF effective Hamiltonian based on only a valence Fock representation~\cite{Xu:2021wwj} and the NNPDF3.1~\cite{NNPDF:2017mvq} and MMHT~\cite{Harland-Lang:2014zoa} global fits. (The inset) the ratio of the valence quark PDFs is compared with the extracted data from JLab MARATHON experiment~\cite{JeffersonLabHallATritium:2021usd}. }
\label{fig_pdf}
\end{center}
\end{figure}
\begin{figure}
\begin{center}
\includegraphics[width=0.95\linewidth]{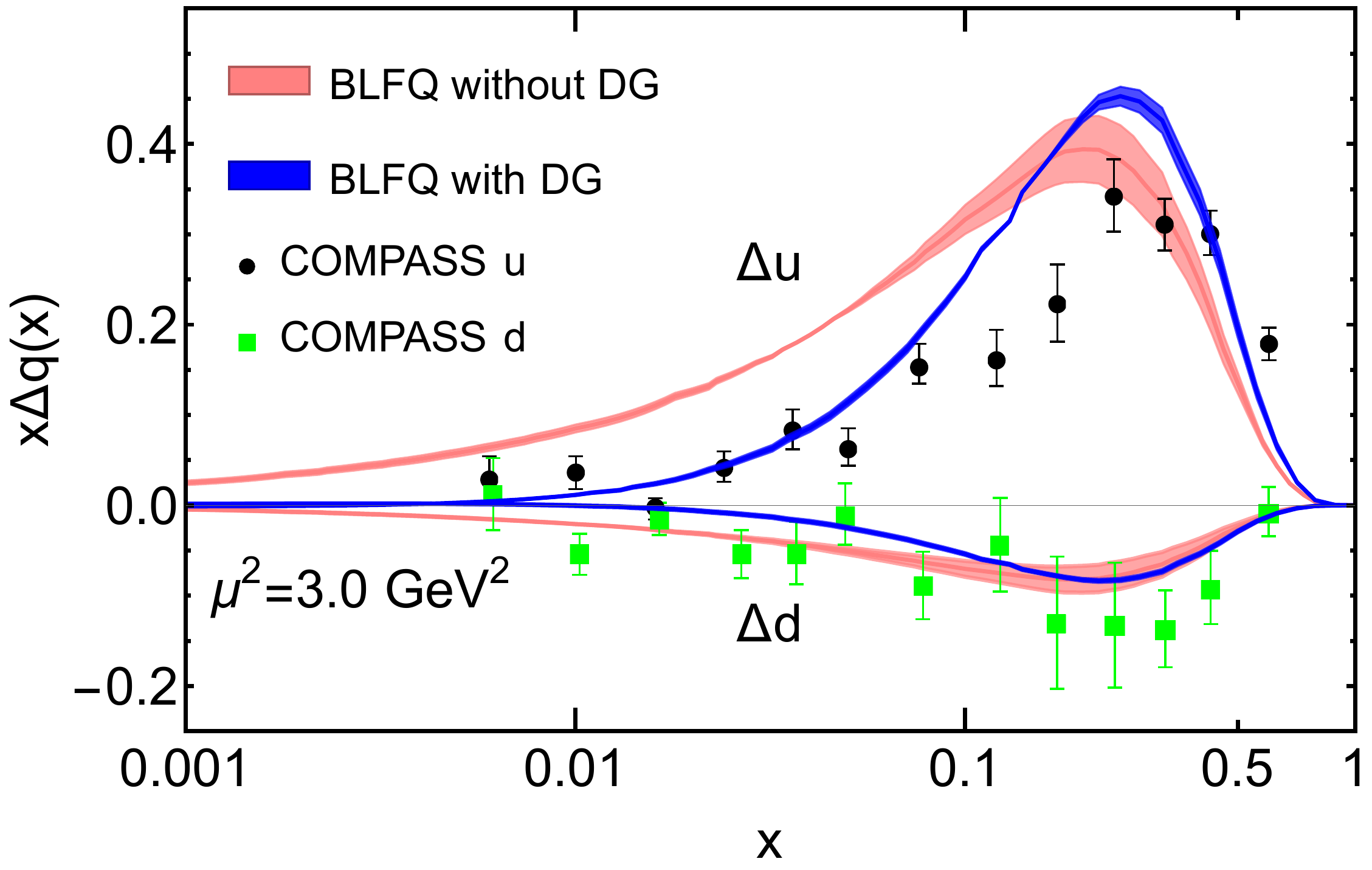}
\caption{The helicity PDFs for the up and down quarks in the proton. We compare BLFQ predictions obtained with one dynamical gluon (blue bands) with the results  previously obtained from a LF effective Hamiltonian based on only a valence Fock representation (pink bands)~\cite{Xu:2021wwj} and the measured data from COMPASS Collaboration~\cite{COMPASS:2010hwr}.}
\label{fig_hpdf}
\end{center}
\end{figure}

To evolve our PDFs from our model scale ($\mu_0^2$) to a higher scale ($\mu^2$), we numerically solve the NNLO DGLAP equations~\cite{Dokshitzer:1977sg,Gribov:1972ri,Altarelli:1977zs} of QCD using the HOPPET~\cite{Salam:2008qg}.
We determine the initial scale by requiring the result after evolution to generate the total first moments of the valence quarks PDFs from the global QCD analysis, $\langle{x}\rangle_u+\langle{x}\rangle_d=0.3742$ at $10$ GeV$^2$~\cite{deTeramond:2018ecg}. This yields $\mu_0^2=0.23\sim 0.25$ GeV$^2$.
 and we then evolve our initial PDFs to the relevant experimental scales.

Figure~\ref{fig_pdf} shows our results for the proton unpolarized PDFs at $\mu^2=10$ GeV$^2$, where we compare the valence quarks and gluon distributions after QCD evolution with the NNPDF3.1~\cite{NNPDF:2017mvq} and MMHT~\cite{Harland-Lang:2014zoa} global fits. A similar comparison can be made with the other global fits of the quark and gluon PDFs~\cite{Hou:2019efy,Moffat:2021dji,Accardi:2016qay,Bailey:2020ooq,H1:2015ubc}.
We also include the proton PDFs previously obtained from a LF effective Hamiltonian~\cite{Xu:2021wwj} based on a valence Fock representation for comparison. 
The error bands in our evolved distributions are reflected from an adopted $10\%$ uncertainty in our model scale. We find a good consistency between our prediction for the proton's valence quark distributions and the global fit. The ratio $d^v(x)/u^v(x)$ is also in reasonable agreement with the extracted data from the MARATHON experiment at JLab~\cite{JeffersonLabHallATritium:2021usd}. A robust method for analysing and extrapolating JLab MARATHON data results in the proton valence-quark ratio: $\lim_{x\to 1} d^v/u^v=0.230\pm 0.057$~\cite{Cui:2021gzg}. We predict the value of $d^v/u^v=0.225\pm 0.025$ at $x\to 1$ which agrees with the extrapolated result. 

According to the Drell-Yan-West relation~\cite{Drell:1969km,West:1970av}, at large $\mu^2$ the valence quark distributions fall off at large $x$ as $(1-x)^p$, where $p$ is associated to the number of valence quarks and for the proton $p=3$. We find that the up quark unpolarized PDF falls off at large $x$ as $(1-x)^{3.2\pm 0.1}$, whereas for the down quark the PDF exhibits $(1-x)^{3.5\pm 0.1}$. Our findings support the perturbative QCD prediction~\cite{Brodsky:1994kg}.  We observe that the gluon PDF is suppressed  at low-$x$ and moves towards the global fits~\cite{NNPDF:2017mvq,Harland-Lang:2014zoa} with the addition of the dynamical gluon, while the distribution for $x > 0.05$ is in reasonable agreement with the global fit. 



\begin{figure}
\begin{center}
\includegraphics[width=0.95\linewidth]{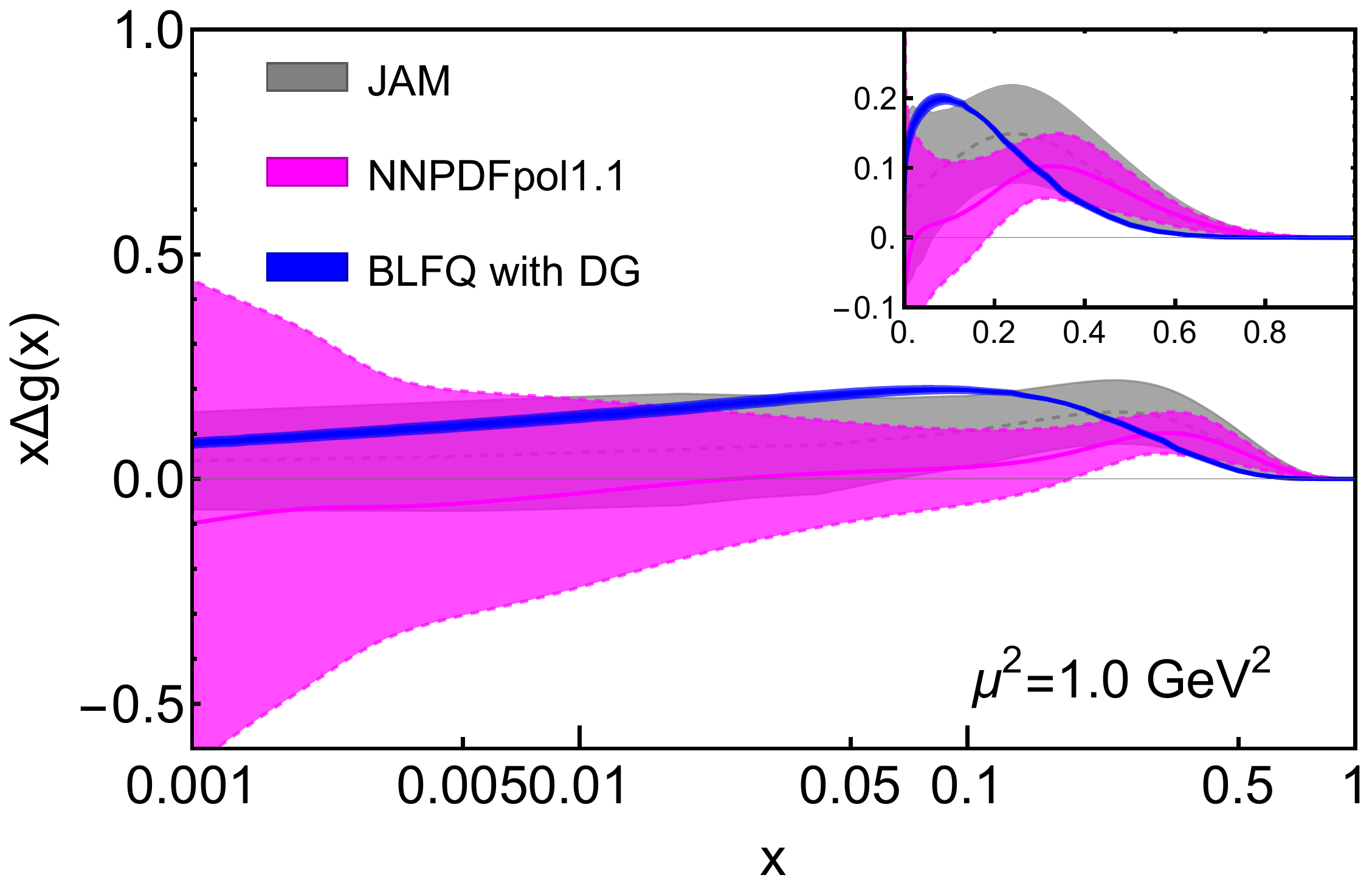}
\caption{The gluon helicity PDF in the proton. We compare our prediction (blue bands) with global analyses by JAM (gray band)~\cite{Sato:2016tuz} and NNPDFpol1.1 (magenta band)~\cite{Nocera:2014gqa}. The inset shows the gluon helicity PDF on a linear scale.}
\label{fig_pdf_gluon}
\end{center}
\end{figure}
\begin{figure}
\begin{center}
\includegraphics[width=0.95\linewidth]{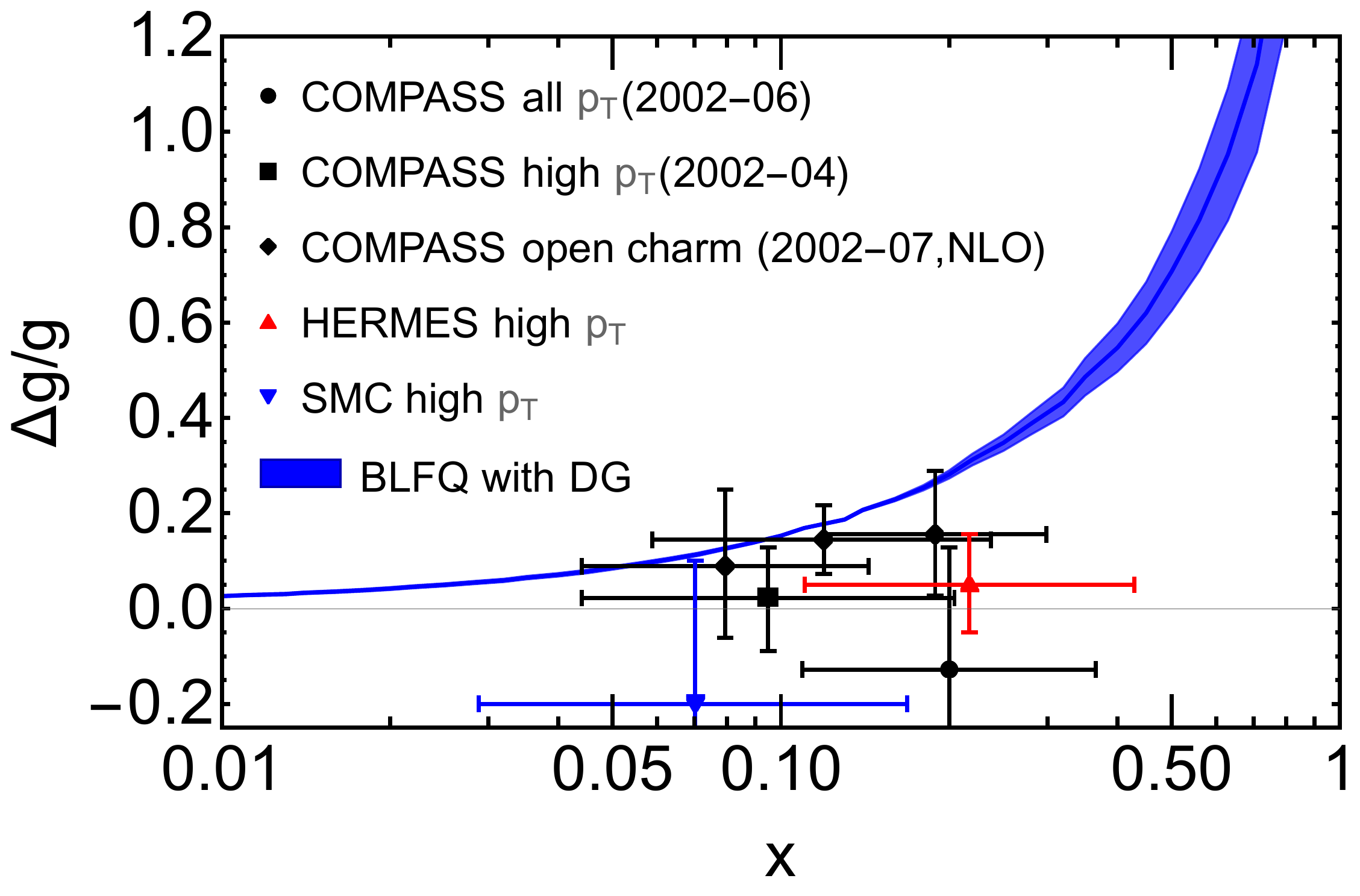}
\caption{The gluon helicity asymmetry, $\Delta g(x)/g(x)$, in the proton. We compare our prediction (blue bands) with the direct measurements of COMPASS~\cite{COMPASS:2005qpp,COMPASS:2015pim}, HERMES~\cite{HERMES:2010nas}, and SMC~\cite{SpinMuonSMC:2004jrx}, which are obtained in LO from high $p_T$ hadrons and from open charm muon production at COMPASS~\cite{COMPASS:2012mpe} in NLO at different $x$.}
\label{fig_asymmetry}
\end{center}
\end{figure}

Figure~\ref{fig_hpdf} shows the helicity PDFs of the up quarks $\Delta u(x)$ and down quarks $\Delta d(x)$, at the scale $\mu^2=3$ GeV$^2$. We find that our  helicity PDFs for both the up and down quarks are reasonably consistent with the experimental data from COMPASS~\cite{COMPASS:2010hwr}. 
For comparison, we also include the quark helicity PDFs previously obtained from a LF effective Hamiltonian based on a valence Fock representation~\cite{Xu:2021wwj}. 
We notice that $\Delta u(x)$ for the up quark improves significantly at small-$x$ region with our current treatment for the nucleon with dynamical gluon.

Utilizing the $|qqqg\rangle$ LFWF amplitudes, we compute the gluon polarized distribution that constitutes the proton spin structure.
We present the gluon helicity PDF at the scale $\mu^2=1$ GeV$^2$ in Fig.~\ref{fig_pdf_gluon}, where we compare our prediction with  the global analyses by the JAM~\cite{Sato:2016tuz} and the NNPDF Collaborations~\cite{Nocera:2014gqa}. 
We observe a fair agreement between our prediction and the global fits at small-$x$, whereas our gluon helicity distribution at large-$x$ falls faster than that of
 the NNPDFpol1.1 analysis. The BLFQ prediction shows somewhat better agreement with the JAM results. Note that there still remain large uncertainties both in the large-x domain and especially in the small-x domain where even the sign is uncertain. ~\cite{Zhou:2022wzm}. 

In Fig.~\ref{fig_asymmetry}, we compare the $\Delta g(x)/g(x)$ ratio obtained from our BLFQ approach with data extracted from high $p_T$ hadrons in the LO analyses~\cite{COMPASS:2005qpp,COMPASS:2015pim} and from the open charm production in the NLO analysis~\cite{COMPASS:2012mpe} at COMPASS, from high $p_T$ hadrons in the LO analyses by the SMC at CERN~\cite{SpinMuonSMC:2004jrx} and at the HERMES experiment~\cite{HERMES:2010nas}. We find a reasonable agreement with the COMPASS data considering the large experimental uncertainties.

The  partonic helicity contributions to the proton spin are given by the first moment of the helicity distributions. Our current analysis yields that, at the model scale, the quark contribution is 
$\frac{1}{2}\Delta\Sigma=0.359\pm 0.002$ to the proton spin, where the up quark contribution, $\frac{1}{2}\Delta\Sigma_u=0.438\pm 0.004$, strongly dominates over the down quark contribution, $\frac{1}{2}\Delta\Sigma_d=-0.080\pm 0.002$. The gluon contribution, $\Delta G=0.131\pm 0.003$, is sizeable to the proton spin.
A recent analysis with updated data sets and PHENIX measurement~\cite{PHENIX:2008swq}  yielded $\Delta G=0.2$ with a constraint: $-0.7 < \Delta G < 0.5$ for  $x_g\in [0.02, 0.3]$. Excluding the $x_g<0.05$ domain, the value of $\Delta G =\int_{0.05}^{0.2} dx~\Delta g(x)=0.23(6)$~\cite{Nocera:2014gqa} and $\Delta G =\int_{0.05}^{1} dx~\Delta g(x)=0.19(6)$~\cite{deFlorian:2014yva} were extracted. The lattice QCD simulations predicted $\Delta G = 0.251(47)(16)$ at the physical pion mass~\cite{Yang:2016plb}. Future  measurements of $\Delta g(x)$ in the $x_g<0.02$ are required to decrease the uncertainty in $\Delta G$. Fortunately,
the upcoming EIC~\cite{Accardi:2012qut,AbdulKhalek:2021gbh} aims to accurately measure the gluon helicity distribution, particularly in the small-$x$ region and provide rigorous limits on the gluon polarized distribution. 


{\it Orbital angular momentum.}---Another important quantity in our understanding of the nucleon spin is the OAM.  The canonical OAM in the light-front gauge is computed using generalized transverse momentum dependent distributions (GTMDs) as~\cite{Lorce:2011kd,Bhattacharya:2022vvo,Liu:2022fvl}
\begin{align}
    {L}^{i}_z=-\int {\rm d} x {\rm d}^2\vec{p}_\perp \frac{\vec{p}_\perp^{~2}}{M^2}F^{i}_{1,4}(x,0,\vec{p}_\perp^{~2},0,0)\,,
\end{align}
with $F^{i}_{1,4}(x,\xi,\vec{p}_\perp^{~2},\vec{p}_\perp\cdot \vec{q}_{\perp},\vec{q}_\perp^{~2})$ being one of the GTMDs for the unpolarized parton~\cite{Meissner:2009ww,Lorce:2013pza}, where $q$ is the momentum transfer and the skewness variable $\xi$ represents the momentum transfer in the longitudinal direction. The GTMD can be expressed in terms of LFWFs as
\begin{align}
  F^i_{1,4} =& \sum_{\Lambda} \int_{\mathcal{N}} \frac{-iM^2}{2(\epsilon^{ij}_{\perp}\vec{p}^i_{\perp}\vec{q}^j_{\perp})}\Lambda\, \Psi^{\Lambda*}_{\{x_i^{\prime},\vec{p}^{\prime}_{i\perp},\lambda_i\}}\Psi^{\Lambda}_{\{x_i,\vec{p}_{i\perp},\lambda_i\}}\nonumber\\ &\times\delta^2(\vec{p}_{\perp}-\frac{\vec{p}^{\prime}_{i\perp}+\vec{p}_{i\perp}}{2})\,\delta(x-x_i)\,,
\end{align}
where $\epsilon^{12}_{\perp}=1$ is used.

We predict that ${L}_z^u=0.0327\pm 0.0013$, ${L}_z^d=-0.0114\pm 0.0004$, and ${L}_z^g=-0.0065 \pm 0.0005$. Note that nothing is known about $L(x)$ experimentally at the present time. Nonetheless, the gluon OAM can be extracted experimentally from the double spin asymmetry in diffractive dijet production~\cite{Bhattacharya:2022vvo}, while the quark OAM can be measured in the exclusive double Drell-Yan process~\cite{Bhattacharya:2017bvs}. Our calculation provides a prediction of the
expected data for the quark and gluon helicities and their OAM from the future experiments as well as baselines for the theoretical
investigations with higher Fock components.

{\it Conclusion and outlook.}---In this letter, we have solved  for the first time the light-front QCD Hamiltonian for the proton within the combined constituent three quarks ($|qqq\rangle$) and three quarks and one gluon ($|qqqg\rangle$) Fock spaces. Together with a three dimensional confinement in the leading Fock sector, the LFWFs obtained as the eigenvectors of this Hamiltonian in Basis Light Front Quantization were employed to compute the the proton initial PDFs. The PDFs at a higher scale have been generated based on the NNLO DGLAP equations and we find reasonable agreement with the global fits for the unpolarized valence quark and gluon distributions. We have predicted that $d^v/u^v=0.225\pm 0.025$ at $x\to 1$ which agrees with the recent analysis from MARATHON experiment yielding $\lim_{x\to 1} d^v/u^v=0.230\pm 0.057$~\cite{Cui:2021gzg}

We have calculated the quark and gluon helicity distributions and their OAM that constitutes the proton spin sum rule. We have observed a good consistency between our predictions for the polarized distributions and the experimental data and/or the global fits. The gluon helicity asymmetry is found to be in fair agreement with the COMPASS data. With one dynamical gluon, we have predicted that the gluon helicity ($\Delta G$) contributes 26$\%$, while the contribution from quark helicity ($\Delta \Sigma$) is $72\%$  to the proton spin. The contributions from the OAM to the proton spin are: ${L}_z^u=0.0327\pm 0.0013$, ${L}_z^d=-0.0114 \pm 0.0004$, and ${L}_z^g=-0.0065\pm 0.0005$. Experimentally, there remain large uncertainties in $\Delta g(x)$, including even the sign, especially in the small-$x$ domain. Future measurements of $\Delta g(x)$ in the $x_g<0.02$ would be most valuable to constrain $\Delta G$. On the other hand, nothing is known about OAM experimentally at the moment. Resolving these issues are major goals of the future EIC~\cite{AbdulKhalek:2021gbh}.

The obtained LFWFs can be further employed to compute the quark and gluon GPDs, TMDs, Wigner distributions as well as the double parton correlations etc., in the nucleon. On the other hand, the present calculation can be straightforwardly extended to higher Fock sectors to incorporate, for example, sea quarks and multi-gluons configurations as well.

\begin{acknowledgments}
{\it Acknowledgments.}---CM is supported by new faculty start up funding by the Institute of Modern Physics, Chinese Academy of Sciences, Grant No. E129952YR0. 
CM also thanks the Chinese Academy of Sciences Presidents International Fellowship Initiative for the support via Grants No. 2021PM0023. XZ is supported by new faculty startup funding by the Institute of Modern Physics, Chinese Academy of Sciences, by Key Research Program of Frontier Sciences, Chinese Academy of Sciences, Grant No. ZDBS-LY-7020, by the Natural Science Foundation of Gansu Province, China, Grant No. 20JR10RA067, by the Foundation for Key Talents of Gansu Province, by the Central Funds Guiding the Local Science and Technology Development of Gansu Province and by the Strategic Priority Research Program of the Chinese Academy of Sciences, Grant No. XDB34000000. YL is supported by the new faculty startup fund of University of Science and Technology of
China. JPV is supported by the Department of Energy under Grants No. DE-FG02-87ER40371, and No. DE-SC0018223 (SciDAC4/NUCLEI). This research used resources of the National Energy Research Scientific Computing Center (NERSC), a U.S. Department of Energy Office of Science User Facility operated under Contract No. DE-AC02-05CH11231.
 A portion of the computational resources were also provided by Gansu Computing Center.
\end{acknowledgments}



\begin{thebibliography}{80}

\bibitem{Callan:1977gz}
C.~G.~Callan, Jr., R.~F.~Dashen and D.~J.~Gross,
``Toward a Theory of the Strong Interactions,''
Phys. Rev. D \textbf{17}, 2717 (1978).

\bibitem{Maris:2003vk}
P.~Maris and C.~D.~Roberts,
``Dyson-Schwinger equations: A Tool for hadron physics,''
Int. J. Mod. Phys. E \textbf{12}, 297-365 (2003).


\bibitem{Roberts:1994dr}
C.~D.~Roberts and A.~G.~Williams,
``Dyson-Schwinger equations and their application to hadronic physics,''
Prog. Part. Nucl. Phys. \textbf{33}, 477-575 (1994).

\bibitem{Bashir:2012fs}
A.~Bashir, L.~Chang, I.~C.~Cloet, B.~El-Bennich, Y.~X.~Liu, C.~D.~Roberts and P.~C.~Tandy,
``Collective perspective on advances in Dyson-Schwinger Equation QCD,''
Commun. Theor. Phys. \textbf{58}, 79-134 (2012).


\bibitem{Joo:2019byq}
B.~Jo\'o \textit{et al.} [USQCD],
``Status and Future Perspectives for Lattice Gauge Theory Calculations to the Exascale and Beyond,''
Eur. Phys. J. A \textbf{55}, no.11, 199 (2019).

\bibitem{Bazavov:2009bb}
A.~Bazavov \textit{et al.} [MILC],
``Nonperturbative QCD Simulations with 2+1 Flavors of Improved Staggered Quarks,''
Rev. Mod. Phys. \textbf{82}, 1349-1417 (2010).


\bibitem{Durr:2008zz}
S.~Durr, Z.~Fodor, J.~Frison, C.~Hoelbling, R.~Hoffmann, S.~D.~Katz, S.~Krieg, T.~Kurth, L.~Lellouch and T.~Lippert, \textit{et al.}
``Ab-Initio Determination of Light Hadron Masses,''
Science \textbf{322}, 1224-1227 (2008).

\bibitem{Hagler:2009ni}
P.~Hagler,
``Hadron structure from lattice quantum chromodynamics,''
Phys. Rept. \textbf{490}, 49-175 (2010).


\bibitem{Brodsky:1997de}
S.~J.~Brodsky, H.~C.~Pauli and S.~S.~Pinsky,
``Quantum chromodynamics and other field theories on the light cone,''
Phys. Rept. \textbf{301}, 299-486 (1998).

\bibitem{Bakker:2013cea}
B.~L.~G.~Bakker, A.~Bassetto, S.~J.~Brodsky, W.~Broniowski, S.~Dalley, \textit{et al.}
``Light-Front Quantum Chromodynamics: A framework for the analysis of hadron physics,''
Nucl. Phys. B Proc. Suppl. \textbf{251-252}, 165-174 (2014).

\bibitem{Brodsky:2014yha}
S.~J.~Brodsky, G.~F.~de Teramond, H.~G.~Dosch and J.~Erlich,
``Light-Front Holographic QCD and Emerging Confinement,''
Phys. Rept. \textbf{584}, 1-105 (2015).


\bibitem{deTeramond:2008ht}
G.~F.~de Teramond and S.~J.~Brodsky,
``Light-Front Holography: A First Approximation to QCD,''
Phys. Rev. Lett. \textbf{102}, 081601 (2009).

\bibitem{deTeramond:2005su}
G.~F.~de Teramond and S.~J.~Brodsky,
``Hadronic spectrum of a holographic dual of QCD,''
Phys. Rev. Lett. \textbf{94}, 201601 (2005).



\bibitem{Vary:2009gt}
J.~P.~Vary, H.~Honkanen, J.~Li, P.~Maris, S.~J.~Brodsky, A.~Harindranath, G.~F.~de Teramond, P.~Sternberg, E.~G.~Ng and C.~Yang,
``Hamiltonian light-front field theory in a basis function approach,''
Phys. Rev. C \textbf{81}, 035205 (2010).

\bibitem{Zhao:2014xaa}
X.~Zhao, H.~Honkanen, P.~Maris, J.~P.~Vary and S.~J.~Brodsky,
``Electron g-2 in Light-Front Quantization,''
Phys. Lett. B \textbf{737}, 65-69 (2014).

\bibitem{Nair:2022evk}
S.~Nair \textit{et al.} [BLFQ],
Phys. Lett. B \textbf{827}, 137005 (2022).

\bibitem{Wiecki:2014ola}
P.~Wiecki, Y.~Li, X.~Zhao, P.~Maris and J.~P.~Vary,
``Basis Light-Front Quantization Approach to Positronium,''
Phys. Rev. D \textbf{91}, no.10, 105009 (2015).


\bibitem{Li:2015zda}
Y.~Li, P.~Maris, X.~Zhao and J.~P.~Vary,
``Heavy Quarkonium in a Holographic Basis,''
Phys. Lett. B \textbf{758}, 118-124 (2016).

\bibitem{Jia:2018ary}
S.~Jia and J.~P.~Vary,
``Basis light front quantization for the charged light mesons with color singlet Nambu\textendash{}Jona-Lasinio interactions,''
Phys. Rev. C \textbf{99}, no.3, 035206 (2019).



\bibitem{Tang:2019gvn}
S.~Tang, Y.~Li, P.~Maris and J.~P.~Vary,
``Heavy-light mesons on the light front,''
Eur. Phys. J. C \textbf{80}, no.6, 522 (2020).

\bibitem{Lan:2019vui}
J.~Lan, C.~Mondal, S.~Jia, X.~Zhao and J.~P.~Vary,
``Parton Distribution Functions from a Light Front Hamiltonian and QCD Evolution for Light Mesons,''
Phys. Rev. Lett. \textbf{122}, no.17, 172001 (2019).

\bibitem{Xu:2019xhk}
C.~Mondal, S.~Xu, J.~Lan, X.~Zhao, Y.~Li, D.~Chakrabarti and J.~P.~Vary,
``Proton structure from a light-front Hamiltonian,''
Phys. Rev. D \textbf{102}, no.1, 016008 (2020).

\bibitem{Xu:2021wwj}
S.~Xu, C.~Mondal, J.~Lan, X.~Zhao, Y.~Li, J.~P.~Vary [BLFQ],
``Nucleon structure from basis light-front quantization,''
Phys. Rev. D \textbf{104}, no.9, 094036 (2021).


\bibitem{Lan:2021wok}
J.~Lan \textit{et al.} [BLFQ],
``Light mesons with one dynamical gluon on the light front,''
Phys. Lett. B \textbf{825}, 136890 (2022).

\bibitem{Kuang:2022vdy}
Z.~Kuang, K.~Serafin, X.~Zhao and J.~P.~Vary,
``All-charm tetraquark in front form dynamics,''
Phys. Rev. D \textbf{105}, 094028 (2022).


\bibitem{STAR:2014wox}
L.~Adamczyk \textit{et al.} [STAR],
``Precision Measurement of the Longitudinal Double-spin Asymmetry for Inclusive Jet Production in Polarized Proton Collisions at $\sqrt{s}=200$ GeV,''
Phys. Rev. Lett. \textbf{115}, 092002  (2015).

\bibitem{deFlorian:2014yva}
D.~de Florian, R.~Sassot, M.~Stratmann and W.~Vogelsang,
``Evidence for polarization of gluons in the proton,''
Phys. Rev. Lett. \textbf{113}, 012001  (2014).

\bibitem{Nocera:2014gqa}
E.~R.~Nocera \textit{et al.} [NNPDF],
``A first unbiased global determination of polarized PDFs and their uncertainties,''
Nucl. Phys. B \textbf{887}, 276-308  (2014).

\bibitem{Ethier:2017zbq}
J.~J.~Ethier, N.~Sato and W.~Melnitchouk,
``First simultaneous extraction of spin-dependent parton distributions and fragmentation functions from a global QCD analysis,''
Phys. Rev. Lett. \textbf{119}, 132001  (2017).

\bibitem{STAR:2021mqa}
M.~S.~Abdallah \textit{et al.} [STAR and (STAR Collaboration)\textdagger{}],
``Longitudinal double-spin asymmetry for inclusive jet and dijet production in polarized proton collisions at $\sqrt{s}=510$ GeV,''
Phys. Rev. D \textbf{105}, no.9, 092011 (2022).

\bibitem{Ji:2020ena}
X.~Ji, F.~Yuan and Y.~Zhao,
``What we know and what we don\textquoteright{}t know about the proton spin after 30 years,''
Nature Rev. Phys. \textbf{3}, no.1, 27-38 (2021).

\bibitem{AbdulKhalek:2021gbh}
R.~Abdul Khalek, A.~Accardi, J.~Adam, D.~Adamiak, W.~Akers, M.~Albaladejo, A.~Al-bataineh, M.~G.~Alexeev, F.~Ameli and P.~Antonioli, \textit{et al.}
``Science Requirements and Detector Concepts for the Electron-Ion Collider: EIC Yellow Report,''
Nucl. Phys. A \textbf{1026}, 122447 (2022).

\bibitem{Thomas:2008ga}
A.~W.~Thomas,
``Interplay of Spin and Orbital Angular Momentum in the Proton,''
Phys. Rev. Lett. \textbf{101}, 102003 (2008).

\bibitem{Jia:2012wf}
S.~Jia and F.~Huang,
``Scale dependencies of proton spin constituents with a nonperturbative alphas,''
Phys. Rev. D \textbf{86}, 094035 (2012).

\bibitem{Zhao:2014hpa}
X.~Zhao,
``Advances in Basis Light-front Quantization,''
Few Body Syst. \textbf{56}, no.6-9, 257-265 (2015).

\bibitem{Zhao:2020kuf}
X.~Zhao, K.~Fu, H.~Zhao and J.~P.~Vary,
``Positronium: an illustration of nonperturbative renormalization in a basis light-front approach,''
PoS \textbf{LC2019}, 090 (2020).

\bibitem{Glazek:1992aq}
S.~D.~Glazek and R.~J.~Perry,
``Special example of relativistic Hamiltonian field theory,''
Phys. Rev. D \textbf{45}, 3740-3754 (1992).

\bibitem{Alkofer:2000wg}
R.~Alkofer and L.~von Smekal,
Phys. Rept. \textbf{353}, 281 (2001)
doi:10.1016/S0370-1573(01)00010-2
[arXiv:hep-ph/0007355 [hep-ph]].

\bibitem{Cates:2011pz} 
  G.~D.~Cates, C.~W.~de Jager, S.~Riordan and B.~Wojtsekhowski,
  ``Flavor decomposition of the elastic nucleon electromagnetic FFs,''
  Phys.\ Rev.\ Lett.\  {\bf 106}, 252003 (2011).

\bibitem{Qattan:2012zf} 
  I.~A.~Qattan and J.~Arrington,
  ``Flavor decomposition of the nucleon electromagnetic FFs,''
  Phys.\ Rev.\ C {\bf 86}, 065210 (2012).
  
  \bibitem{Diehl:2013xca} 
  M.~Diehl and P.~Kroll,
  ``Nucleon FFs, generalized parton distributions and quark angular momentum,''
  Eur.\ Phys.\ J.\ C {\bf 73}, no. 4, 2397 (2013).
  
\bibitem{Ernst:1960zza}
F.~J.~Ernst, R.~G.~Sachs and K.~C.~Wali,
``Electromagnetic form factors of the nucleon,''
Phys. Rev. \textbf{119}, 1105-1114 (1960).
  
%

\bibitem{Lin:2021xrc}
Y.~H.~Lin, H.~W.~Hammer and U.~G.~Mei\ss{}ner,
``New Insights into the Nucleon\textquoteright{}s Electromagnetic Structure,''
Phys. Rev. Lett. \textbf{128}, no.5, 052002 (2022).

\bibitem{Alexandrou:2018sjm}
C.~Alexandrou, S.~Bacchio, M.~Constantinou, J.~Finkenrath, K.~Hadjiyiannakou, K.~Jansen, G.~Koutsou and A.~Vaquero Aviles-Casco,
``Proton and neutron electromagnetic form factors from lattice QCD,''
Phys. Rev. D \textbf{100}, no.1, 014509 (2019).



\bibitem{Tanabashi:2018oca}
M.~Tanabashi \textit{et al.} [Particle Data Group],
``Review of Particle Physics,''
Phys. Rev. D \textbf{98}, no.3, 030001 (2018).
6605 citations counted in INSPIRE as of 30 Mar 2021




\bibitem{Dokshitzer:1977sg}
  Y.~L.~Dokshitzer,
  ``Calculation of the Structure Functions for Deep Inelastic Scattering and e+ e- Annihilation by Perturbation Theory in Quantum Chromodynamics.,''
  Sov.\ Phys.\ JETP {\bf 46}, 641 (1977)
  [Zh.\ Eksp.\ Teor.\ Fiz.\  {\bf 73}, 1216 (1977)].
  
  \bibitem{Gribov:1972ri} 
  V.~N.~Gribov and L.~N.~Lipatov,
  ``Deep inelastic e p scattering in perturbation theory,''
  Sov.\ J.\ Nucl.\ Phys.\  {\bf 15}, 438 (1972).
  
  \bibitem{Altarelli:1977zs} 
  G.~Altarelli and G.~Parisi,
  ``Asymptotic Freedom in Parton Language,''
  Nucl.\ Phys.\ B {\bf 126}, 298 (1977).


\bibitem{Salam:2008qg} 
  G.~P.~Salam and J.~Rojo,
  ``A Higher Order Perturbative Parton Evolution Toolkit (HOPPET),''
  Comput.\ Phys.\ Commun.\  {\bf 180}, 120 (2009).
  
\bibitem{deTeramond:2018ecg}
G.~F.~de Teramond \textit{et al.} [HLFHS],
``Universality of Generalized Parton Distributions in Light-Front Holographic QCD,''
Phys. Rev. Lett. \textbf{120}, no.18, 182001 (2018).

\bibitem{NNPDF:2017mvq}
R.~D.~Ball \textit{et al.} [NNPDF],
``Parton distributions from high-precision collider data,''
Eur. Phys. J. C \textbf{77}, no.10, 663 (2017).

\bibitem{Harland-Lang:2014zoa}
L.~A.~Harland-Lang, A.~D.~Martin, P.~Motylinski and R.~S.~Thorne,
``Parton distributions in the LHC era: MMHT 2014 PDFs,''
Eur. Phys. J. C \textbf{75}, no.5, 204 (2015).



%
%
%
%
%
%
%
%
%
%
%
%



\bibitem{COMPASS:2010hwr}
M.~G.~Alekseev \textit{et al.} [COMPASS],
``Quark helicity distributions from longitudinal spin asymmetries in muon-proton and muon-deuteron scattering,''
Phys. Lett. B \textbf{693}, 227-235 (2010).


\bibitem{Hou:2019efy}
T.~J.~Hou, J.~Gao, T.~J.~Hobbs, K.~Xie, S.~Dulat, M.~Guzzi, J.~Huston, P.~Nadolsky, J.~Pumplin and C.~Schmidt, \textit{et al.}
``New CTEQ global analysis of quantum chromodynamics with high-precision data from the LHC,''
Phys. Rev. D \textbf{103}, no.1, 014013 (2021).

\bibitem{Moffat:2021dji}
E.~Moffat \textit{et al.} [Jefferson Lab Angular Momentum (JAM)],
``Simultaneous Monte~Carlo analysis of parton densities and fragmentation functions,''
Phys. Rev. D \textbf{104}, no.1, 016015 (2021).

\bibitem{Accardi:2016qay}
A.~Accardi, L.~T.~Brady, W.~Melnitchouk, J.~F.~Owens and N.~Sato,
``Constraints on large-$x$ parton distributions from new weak boson production and deep-inelastic scattering data,''
Phys. Rev. D \textbf{93}, no.11, 114017 (2016).

\bibitem{Bailey:2020ooq}
S.~Bailey, T.~Cridge, L.~A.~Harland-Lang, A.~D.~Martin and R.~S.~Thorne,
``Parton distributions from LHC, HERA, Tevatron and fixed target data: MSHT20 PDFs,''
Eur. Phys. J. C \textbf{81}, no.4, 341 (2021).

\bibitem{H1:2015ubc}
H.~Abramowicz \textit{et al.} [H1 and ZEUS],
``Combination of measurements of inclusive deep inelastic ${e^{\pm }p}$ scattering cross sections and QCD analysis of HERA data,''
Eur. Phys. J. C \textbf{75}, no.12, 580 (2015).

\bibitem{Drell:1969km}
S.~D.~Drell and T.~M.~Yan,
``Connection of Elastic Electromagnetic Nucleon Form-Factors at Large Q**2 and Deep Inelastic Structure Functions Near Threshold,''
Phys. Rev. Lett. \textbf{24}, 181-185 (1970).
  
\bibitem{West:1970av}
G.~B.~West,
``Phenomenological model for the electromagnetic structure of the proton,''
Phys. Rev. Lett. \textbf{24}, 1206-1209 (1970).
  
\bibitem{Brodsky:1994kg}
S.~J.~Brodsky, M.~Burkardt and I.~Schmidt,
``Perturbative QCD constraints on the shape of polarized quark and gluon distributions,''
Nucl. Phys. B \textbf{441}, 197-214 (1995).

\bibitem{JeffersonLabHallATritium:2021usd}
D.~Abrams \textit{et al.} [Jefferson Lab Hall A Tritium],
``Measurement of the Nucleon $F^n_2/F^p_2$ Structure Function Ratio by the Jefferson Lab MARATHON Tritium/Helium-3 Deep Inelastic Scattering Experiment,''
Phys. Rev. Lett. \textbf{128}, no.13, 132003 (2022).

\bibitem{Cui:2021gzg}
Z.~F.~Cui, F.~Gao, D.~Binosi, L.~Chang, C.~D.~Roberts and S.~M.~Schmidt,
``Valence Quark Ratio in the Proton,''
Chin. Phys. Lett. \textbf{39}, no.4, 041401 (2022).

\bibitem{Sato:2016tuz}
N.~Sato \textit{et al.} [Jefferson Lab Angular Momentum],
``Iterative Monte Carlo analysis of spin-dependent parton distributions,''
Phys. Rev. D \textbf{93}, no.7, 074005 (2016).

\bibitem{Zhou:2022wzm}
Y.~Zhou \textit{et al.} [Jefferson Lab Angular Momentum (JAM)],
``How well do we know the gluon polarization in the proton?,''
Phys. Rev. D \textbf{105}, no.7, 074022 (2022).


\bibitem{COMPASS:2005qpp}
E.~S.~Ageev \textit{et al.} [COMPASS],
``Gluon polarization in the nucleon from quasi-real photoproduction of high-p(T) hadron pairs,''
Phys. Lett. B \textbf{633}, 25-32 (2006).

\bibitem{COMPASS:2015pim}
C.~Adolph \textit{et al.} [COMPASS],
``Leading-order determination of the gluon polarisation from semi-inclusive deep inelastic scattering data,''
Eur. Phys. J. C \textbf{77}, no.4, 209 (2017).

\bibitem{COMPASS:2012mpe}
C.~Adolph \textit{et al.} [COMPASS],
``Leading and Next-to-Leading Order Gluon Polarization in the Nucleon and Longitudinal Double Spin Asymmetries from Open Charm Muoproduction,''
Phys. Rev. D \textbf{87}, no.5, 052018 (2013).

\bibitem{SpinMuonSMC:2004jrx}
B.~Adeva \textit{et al.} [Spin Muon (SMC)],
``Spin asymmetries for events with high p(T) hadrons in DIS and an evaluation of the gluon polarization,''
Phys. Rev. D \textbf{70}, 012002 (2004).

\bibitem{HERMES:2010nas}
A.~Airapetian \textit{et al.} [HERMES],
``Leading-Order Determination of the Gluon Polarization from high-p(T) Hadron Electroproduction,''
JHEP \textbf{08}, 130 (2010).

\bibitem{PHENIX:2008swq}
A.~Adare \textit{et al.} [PHENIX],
``The Polarized gluon contribution to the proton spin from the double helicity asymmetry in inclusive pi0 production in polarized p + p collisions at s**(1/2) = 200-GeV,''
Phys. Rev. Lett. \textbf{103}, 012003 (2009).

\bibitem{Yang:2016plb}
Y.~B.~Yang, R.~S.~Sufian, A.~Alexandru, T.~Draper, M.~J.~Glatzmaier, K.~F.~Liu and Y.~Zhao,
``Glue Spin and Helicity in the Proton from Lattice QCD,''
Phys. Rev. Lett. \textbf{118}, no.10, 102001 (2017).

\bibitem{Accardi:2012qut}
A.~Accardi, J.~L.~Albacete, M.~Anselmino, N.~Armesto, E.~C.~Aschenauer, A.~Bacchetta, D.~Boer, W.~K.~Brooks, T.~Burton and N.~B.~Chang, \textit{et al.}
``Electron Ion Collider: The Next QCD Frontier: Understanding the glue that binds us all,''
Eur. Phys. J. A \textbf{52}, no.9, 268 (2016).


\bibitem{Lorce:2011kd}
C.~Lorce and B.~Pasquini,
``Quark Wigner Distributions and Orbital Angular Momentum,''
Phys. Rev. D \textbf{84}, 014015 (2011).

\bibitem{Bhattacharya:2022vvo}
S.~Bhattacharya, R.~Boussarie and Y.~Hatta,
``Signature of the Gluon Orbital Angular Momentum,''
Phys. Rev. Lett. \textbf{128}, no.18, 182002 (2022).

\bibitem{Liu:2022fvl}
Y.~Liu \textit{et al.} [BLFQ],
Phys. Rev. D \textbf{105}, no.9, 094018 (2022)
doi:10.1103/PhysRevD.105.094018
[arXiv:2202.00985 [hep-ph]].

\bibitem{Meissner:2009ww}
S.~Meissner, A.~Metz and M.~Schlegel,
``Generalized parton correlation functions for a spin-1/2 hadron,''
JHEP \textbf{08}, 056 (2009).


\bibitem{Lorce:2013pza}
C.~Lorc\'e and B.~Pasquini,
``Structure analysis of the generalized correlator of quark and gluon for a spin-1/2 target,''
JHEP \textbf{09}, 138 (2013).

\bibitem{Bhattacharya:2017bvs}
S.~Bhattacharya, A.~Metz and J.~Zhou,
``Generalized TMDs and the exclusive double Drell\textendash{}Yan process,''
Phys. Lett. B \textbf{771}, 396-400 (2017)
[erratum: Phys. Lett. B \textbf{810}, 135866 (2020)].





\end{thebibliography}
\end{document}